\documentclass[manuscript]{aastex}
\usepackage{tikz}
\usepackage{amsmath}

\usepackage{graphicx}

\usepackage{pdflscape}

\slugcomment{\textbf{Revised 2024 December 23} }

\shorttitle{Kreutz C/2024 S1}
\shortauthors{Jewitt}

\begin{document}

\title{Demise of Kreutz Sungrazing Comet C/2024 S1 (ATLAS)}

\author{
David Jewitt$^{1}$, 
Jane Luu$^{2}$ and
Jing Li$^{1}$
} 
\affil{$^1$Department of Earth, Planetary and Space Sciences, UCLA, \\595 Charles Young Drive, Los Angeles, CA 90095\\
$^2$HAB, Department of Geosciences, University of Oslo \\
P.O. Box 1047, Blindern, NO-0316 Oslo, Norway}

\email{djewitt@gmail.com}

\begin{abstract}
Most Kreutz family sungrazing comets are discovered only days before perihelion, severely limiting observational opportunities to study their physical nature and decay.  Kreutz sungrazer C/2024 S1 (ATLAS) was discovered a month before reaching its perihelion distance of 0.008 au, allowing physical observations from both ground- and space-based telescopes.  We present observations from 0.9 au to 0.4 au using the Nordic Optical Telescope  showing that 1) nucleus disintegration was on-going already at 0.7 au pre-perihelion,  2) the activity varied unpredictably with distance and 3) the  nucleus radius was $<$600 m (red geometric albedo 0.04 assumed).  We also use coronagraphic observations from the STEREO-A spacecraft to study C/2024 S1 at heliocentric distances $\lesssim$0.1 au.  We find that the coma scattering cross-section peaked near 0.075 au and faded progressively, by a factor $\sim$20, towards the last observation at 0.02 au.  We interpret the near-perihelion fading as a result of the sublimation of refractory coma grains, beginning at  blackbody temperatures $\sim$1000 K, consistent with olivine composition.  The comet was not detected after perihelion.   We consider processes operating to destroy the nucleus when near perihelion, concluding that  rotational instability and sublimation losses work together towards this end, even before entry of the comet into the Roche lobe of the Sun.

\end{abstract}


\section{INTRODUCTION}
\label{intro}

Comet C/2024 S1 (ATLAS) (hereafter S1) was discovered on UT 2024 September 27 by R. Siverd using data from the ATLAS survey telescope \citep{Siv24}.  Comet S1 had semimajor axis $a$ = 96.6 au, inclination $i$ = 141.9\degr, and eccentricity $e$ = 0.99992.  The perihelion distance, $q = 0.0080$ au, corresponds to 1.73 R$_{\odot}$, where R$_{\odot} = 6.96\times10^8$ m is the radius of the Sun.  Both the small perihelion and the large inclination suggest that C/2024 S1 is a member of the Kreutz sungrazing comet group \citep{Mar67}. This association is cemented by the argument of perihelion, $\omega$ = 69.2\degr, and the longitude of the ascending node, $\Omega$ = 347.1\degr, both of which are Kreutz-like.  Kreutz comets are fragments of a precursor body, likely from the Oort cloud, that broke up as a result of a previous close approach to the Sun (\cite{Sek07}, \cite{Fer21}).  The physical lifetimes of the Kreutz comets are limited by intense heating when near the Sun.  Individual Kreutz comet lifetime estimates are $\sim10^3$ year (i.e.~comparable to the 950 year orbit period of C/2024 S1), consistent with the observation that a majority of such objects do not appear to survive perihelion \citep{Kni10}.  Modern-day Kreutz comets are likely to be products of cascading fragmentation \citep{Sek07} and the time elapsed since breakup of the original Kreutz precursor nucleus is uncertain, but possibly much greater than 10$^3$ years.  

Most Kreutz comets are  discovered when very close to the Sun, in data obtained by orbiting solar coronagraphic observatories (principally the SOHO and STEREO spacecraft).  These sun-focused spacecraft have very large pixels and, in addition to poor angular resolution, their sensitivity is limited by the high and  variable surface brightness of the solar corona background.  The early discovery of S1 against dark sky in ground-based data, when still 1 au from the Sun and $\sim$1 month before the UT 2024 October 28 perihelion, provided us with the opportunity to acquire higher resolution, more sensitive observations with a ground-based, nighttime telescope. Once lost from the ground in the glare from the Sun, we continued observations using the STEREO Cor-2A space-born coronagraph.  (Contemporaneous LASCO coronagraphic observations from the SOHO spacecraft were publicly available only in ``quick-look'' (uncalibrated) form at the time of writing.  An examination of uncalibrated LASCO images showed the comet to be saturated near perihelion, so we did not use these data in the present study). 
Our scientific objective was to examine the evolution and decay of a sungrazing comet in unprecedented detail. 

\section{OBSERVATIONS}

\textbf{NOT Data:}  We imaged S1 using the 2.56 m diameter Nordic Optical Telescope (NOT) located at 2.4 km altitude in Las Palma, Spain. This active optics telescope can point  to within 10\degr~of the horizon and so is of unusual value in the twilight study of objects at relatively small solar elongations.  We used the ALFOSC (Alhambra Faint Object Spectrograph and Camera) imager, which provides an image scale of 0.21\arcsec~pixel$^{-1}$ over an approximately 6.5\arcmin$\times$6.5\arcmin~unvignetted field of view.  
The images were taken using a Broadband Bessel R filter (central wavelength $\lambda_c$ = 6500\AA, FWHM = 1300\AA) for photometric monitoring purposes.  The pre-perihelion observational window was limited to a maximum of about 20 minutes per night by the altitude limit at the beginning of each session, and by the onset of twilight at the end.  We obtained 6 to 8 images each of 60 s duration in this time, reduced to only two images of 30 s by October 20.  The images were tracked at non-sidereal rates (e.g.~$\sim$200\arcsec~hour$^{-1}$ on October 3) to follow the motion of the comet, and dithered between exposures to provide protection from bad pixels.  Prior to UT October 19, we used field stars and an autoguider to control the non-sidereal tracking of the telescope.  On and after that date, we abandoned the autoguider (which worked poorly in ever brighter skies) and acquired unguided images at non-sidereal rates.  Photometric calibration used images of Landolt standard stars supplemented by field stars observed in the Pan-STARRS photometric catalog.  Typical seeing ranged from $\sim$1\arcsec~FWHM to 5\arcsec; we rejected the worst seeing data in our analysis.  Flat field exposures were obtained each night using an illuminated patch on the inside of the NOT dome.  A journal of NOT observations is given in Table \ref{geometry}.

Photometry was obtained within circular projected apertures of fixed linear (not angular) scale.  Keeping in mind the variable seeing and the angular scale of the coma, we elected to measure  apertures of  radii 5000 km, 10,000 km and 15,000 km projected to the distance of the comet, referred to as $R_5$, $R_{10}$ and $R_{15}$, respectively.  Sky subtraction was obtained from the median signal within concentric annuli large compared to the photometry annuli.  The comet tail crosses the sky annulus as do the trails of some field stars; we checked to be sure that these signals are all effectively removed by the use of the median within the sky region.

\begin{figure}
\epsscale{0.99}
\plotone{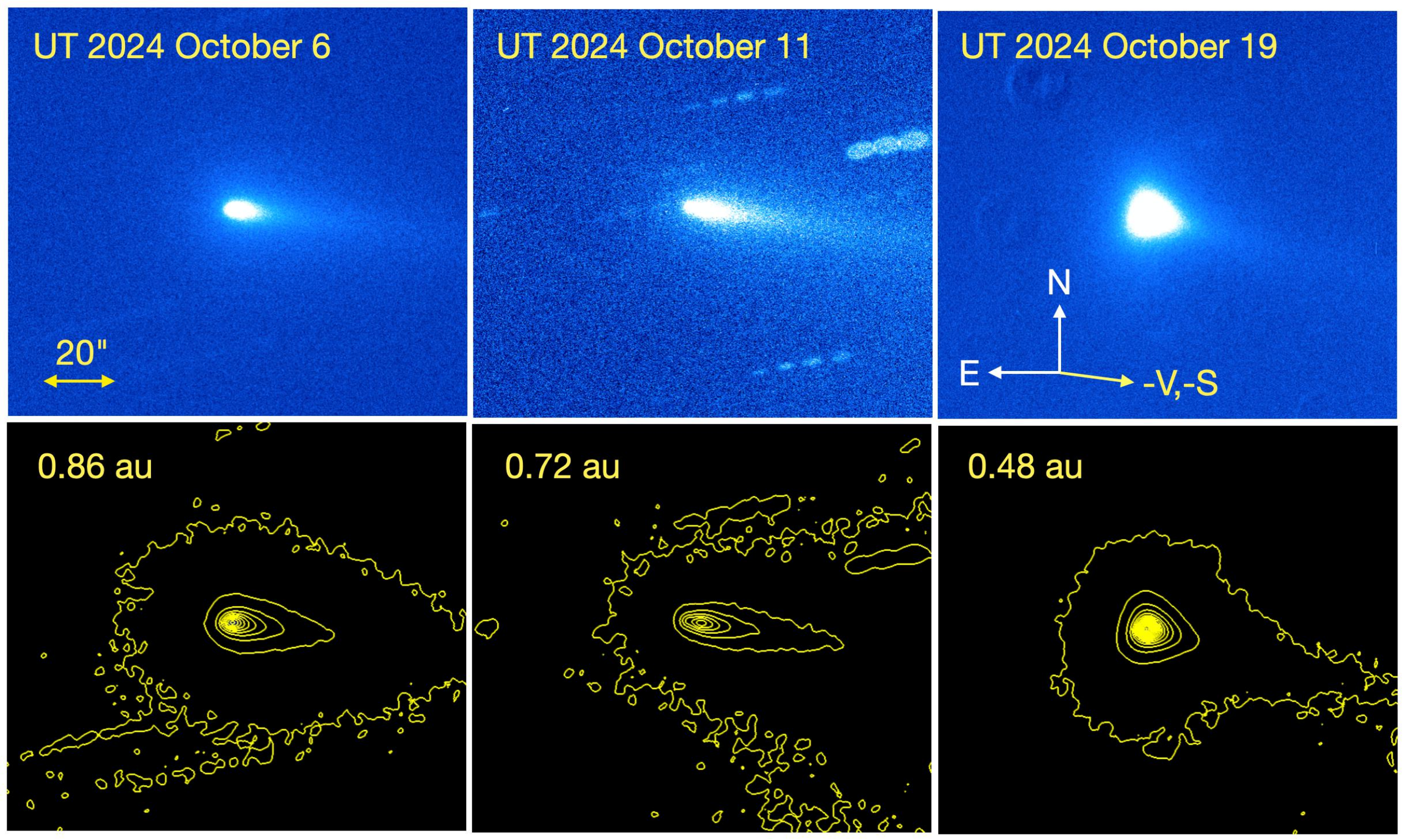}
\caption{Pre-perihelion appearance of C/2024 S1 from NOT images on three dates, as shown.  The scale bar and direction arrows are common to all six panels. \label{NOT}}
\end{figure}

\begin{figure}
\epsscale{0.9}
\plotone{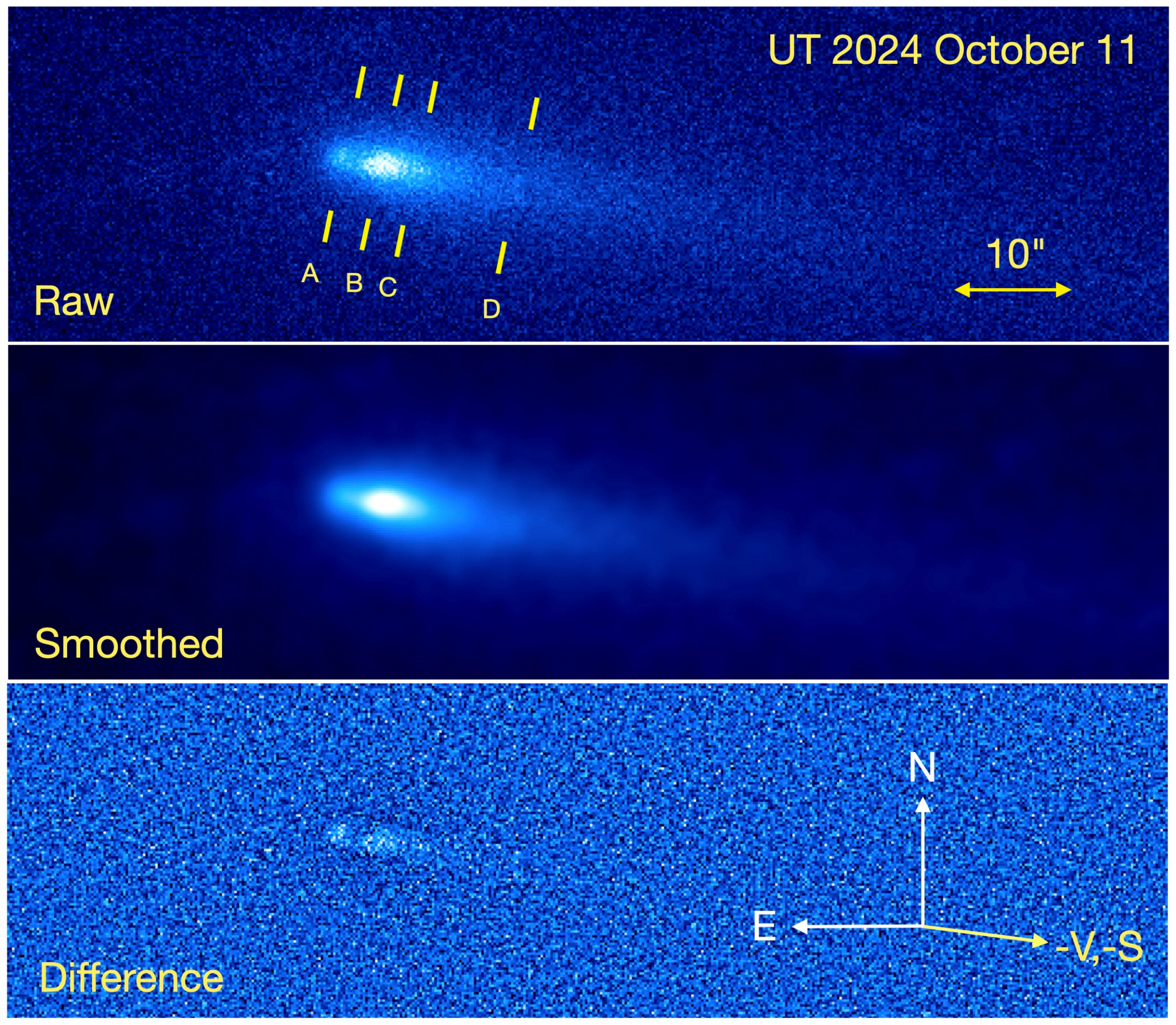}

\caption{NOT image showing the pre-perihelion appearance of C/2024 S1 on UT 2024 October 11 at $r_H$ = 0.725 au.  The image at the top shows raw data compiled from the median of five images each of 60 s integration.  A gaussian smoothed version of this image is shown in the middle panel while the difference image is at the bottom.  Four discrete objects are labeled.  A scale bar and cardinal directions are marked. \label{october11}}
\end{figure}

We show sample composite images in Figure \ref{NOT}.   The early observations show little evidence for morphological or photometric change other than night-to-night differences caused by the variable seeing and sky brightness. The October 6 image in Figure \ref{NOT} is representative in showing a morphology that is typical of a radiation pressure swept comet.  By UT 2024 October 11, clear evidence for breakup is evident in the presence of discrete structures along the tail axis.  This is seen as an elongation of the isophotes in Figure \ref{NOT} and emphasized by a more detailed view in Figure \ref{october11}, where at least four components are evident.  These discrete structures had disappeared by UT October 17, and the apparent brightness by then had increased by $\sim$1.5 magnitudes, presumably related to the exposure of fresh ice by breakup.  Observations on UT October 19 show the comet to have  brightened again to $R_{15}$ = 9.2, some 5.5 magnitudes (a factor of 160) brighter than the plateau level in the October 4 - 11 period.  The brightening partly reflects increased dust production and was accompanied by a marked morphological change (Figure \ref{NOT}), from a spherical head region with an elongated tail to a more blunt sun-facing coma with ``wings'' perpendicular to the Sun - comet line.  Similar winged morphology has been observed in other comets (notably the sungrazer C/2012 S1 (ISON)) and, while the cause is not clear, may relate to an interaction between multiple sources in close proximity \citep{Sam15}.

The spectra of comets with $r_H \lesssim$ 2 au consist of both continuum due to sunlight scattered from the nucleus and from coma dust, and resonance fluorescence emission from molecules in the gas coma.  The photodestruction lifetimes of cometary molecules vary as $r_H^2$ and are very short in near-Sun comets like C/2024 S1.  In addition, gas emission bands from cometary species (e.g., CN, C$_2$, C$_3$) are concentrated towards short optical wavelengths \citep{Her50} and therefore primarily contaminate the B and V broadband filters.  Here, we analyze only R filter photometry, in which we presume that the bulk of the emission is sunlight scattered from cometary dust.  

The absolute magnitude is the magnitude of the comet reduced to $r_H = \Delta$ = 1 au and $\alpha$ = 0\degr.  For  fixed linear aperture photometry it is related to the apparent magnitude by

\begin{equation}
H_R = m_R - 5\log_{10}(r_H\Delta) + 2.5\log_{10}(\Phi(\alpha))
\end{equation}

\noindent where $\Phi({\alpha})$ is the phase function corresponding to phase angle $\alpha$ (normalized to $\Phi(0)$ =1).  The phase angles of our NOT observations were relatively large (51\degr~to 65\degr: Table \ref{geometry}) requiring consideration of the (unmeasured) phase function.  For convenience, we used the tabulation by Schleicher\footnote{\url{https://asteroid.lowell.edu/comet/dustphase/}} with  results that are not substantially different from the Henyey-Greenstein parametrization advocated by \cite{Mar86}.  Fortunately, the strongest angle dependence and the largest uncertainties in $\Phi(\alpha)$ are confined to forward scattering geometries ($\alpha \gtrsim 120\degr$) and the sensitivity to $\alpha$ in the range of the NOT data is modest.  For example, the Schleicher curve gives only a small variation from $\Phi(50\degr)$ = 0.38 to $\Phi(90\degr)$ = 0.54.  Still, we are mindful that the phase function of S1 is an assumed quantity, not a measured one, and expect that it might easily be in error by a factor of order two.

\begin{figure}
\epsscale{0.8}
\plotone{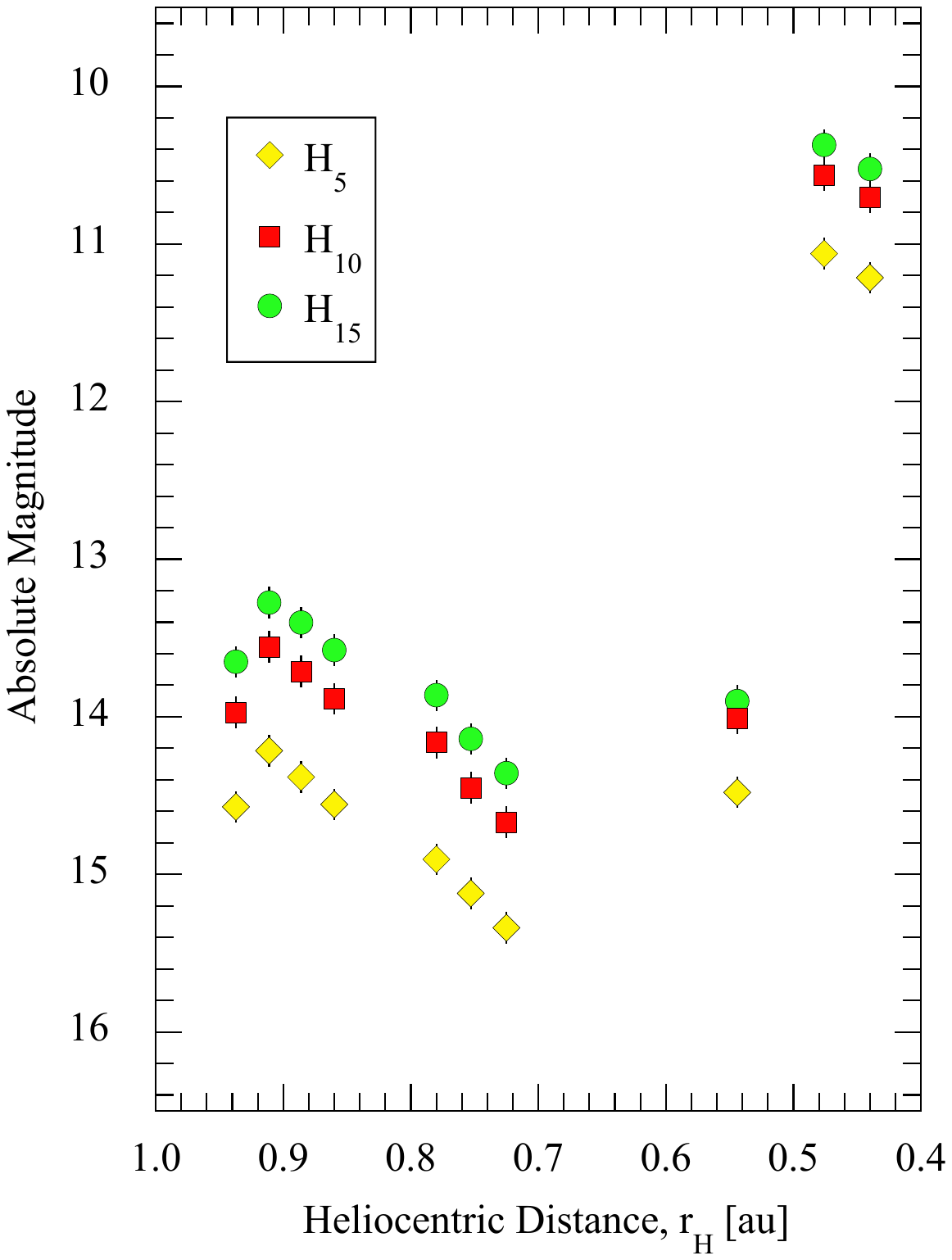}
\caption{Absolute magnitude vs.~heliocentric distance for pre-perihelion NOT observations, measured within circular apertures of fixed projected radii 5,000 km (yellow diamonds), 10,000 km (red squares) and 15,000 km (green circles). \label{NOT_plot}}
\end{figure}

Figure \ref{NOT_plot} shows the NOT absolute magnitudes.  The  magnitude variation cannot be represented by a simple power law function of the heliocentric distance, as is widely done for comets.  Instead, $H$ at first fades in each aperture by $\sim$1 magnitude from  UT October 3 ($r_H$ = 0.937 au) to October 11 (0.725 au) but then reverses its decline by October 17 (0.544 au) and brightens by $\sim$3.5 magnitudes by October 19 (0.476 au).  The aperture differences, $\Delta(5-10) = R_5 - R_{10}$, and $\Delta(10-15) = R_{10} - R_{15}$ and $\Delta(5-15) = R_5 - R_{15}$, provide a crude measure of the steepness of the coma surface brightness  measured as a function of distance from the nucleus, with smaller values corresponding to a steeper surface brightness profile.  Figure \ref{NOT_plot} shows that all three differences shrink on October 17 and 19, consistent with a surface brightness profile steepened by the ejection of fresh material from the nucleus.

The R-band absolute magnitude of the comet is related to the scattering cross-section, $C$ by

\begin{equation}
C \textrm{ [km}^2\textrm{] }= \frac{9.8\times10^5 10^{-0.4 H}}{p_R}
\label{cross}
\end{equation}

\noindent where $p_{R}$ is the R-band geometric albedo and $H$ is the absolute magnitude.  The geometric albedo of S1 is unmeasured so we assume $p_R$ = 0.04 as is found in other comets. The magnitudes $m_R$, $H$ and cross-section $C$ are tabulated in Table \ref{NOT_photometry}.

The large and changing cross-sections in Table \ref{NOT_photometry} reflect scattering from cometary dust in the coma, not the nucleus.  The most stringent limit to the pre-perihelion nucleus is  set by the smallest cross-section, $C_5$ = 17 km$^2$, measured on UT 2024 October 11.  The radius of an equal area circle is  $r_n = (C_5/\pi)^{1/2} \sim$  2.3 km which sets a strong upper limit to the pre-perihelion radius of the nucleus because of dust contamination of the photometry.

To attempt to place a more stringent limit on the pre-perihelion nucleus radius we examined fragment photometry from Oct 11 (Figure \ref{october11}).  On this date, the comet shows at least four discrete sources embedded in the dust tail.  We assume that the eastern-most of these, Component A, is the primary nucleus and that the other sources are fragments pushed away from the Sun by outgassing forces.  That Component B is brighter than A is not surprising, since the brightness of each component is dominated by scattering from dust.  Using a 5 pixel (1.1\arcsec) radius aperture, we find apparent magnitude of Component A as $m_5$ = 19.0, and absolute magnitude $H_5$ = 18.4.  The uncertainty on these numbers is set primarily by systematic errors (e.g., caused by the small aperture, and by the unmeasured phase function).  The cross-section from Equation \ref{cross} is $C$ = 1.1 km$^2$, equal to the area of a circle of radius $r_n$ = 600 m.  Strictly, this is still an upper limit to the radius of the nucleus because Component A is clearly non-stellar.  

We obtained additional NOT images on UT 2024 November 6 and 10 in search of possible surviving post-perihelion remnants of S1.  Images taken in twilight show no sign of the comet, with a point-source magnitude limit  $m_R >$ 18.2 on November 6 and $m_R >$ 19.5 on November 10.  The ephemeris  accuracy on this date (from JPL Horizons) was 1$\sigma \sim \pm$12\arcsec, small compared to the 6.5\arcmin~ALFOSC field of view, so that pointing should not be a concern.  To set a limit on the size of any surviving nucleus, we assumed a C-type asteroid phase function, which predicts a dimming by $-2.5\log_{10}(\Phi(\alpha))$ = 2.6 and 2.5 magnitudes on these two dates (c.f., Table \ref{geometry}) relative to $\alpha$ = 0\degr.  The resulting limit to the absolute magnitude of the post-perihelion nucleus is $H_R >$ 17.2 and $H_R >$ 18.1 on UT November 6 and 10, respectively.  Using the more stringent November 10 observation, Equation \ref{cross} gives a limit to the scattering cross-section $C <$ 1.4 km$^2$ (equivalent spherical radius $r_n <$ 0.7 km). This is essentially the same as the limit to the nucleus radius determined from pre-perihelion photometry.  Therefore, the post-perihelion data do not allow us to conclude that the nucleus was destroyed, although the non-detection is consistent with this possibility.

\begin{figure}
\epsscale{0.95}
\plotone{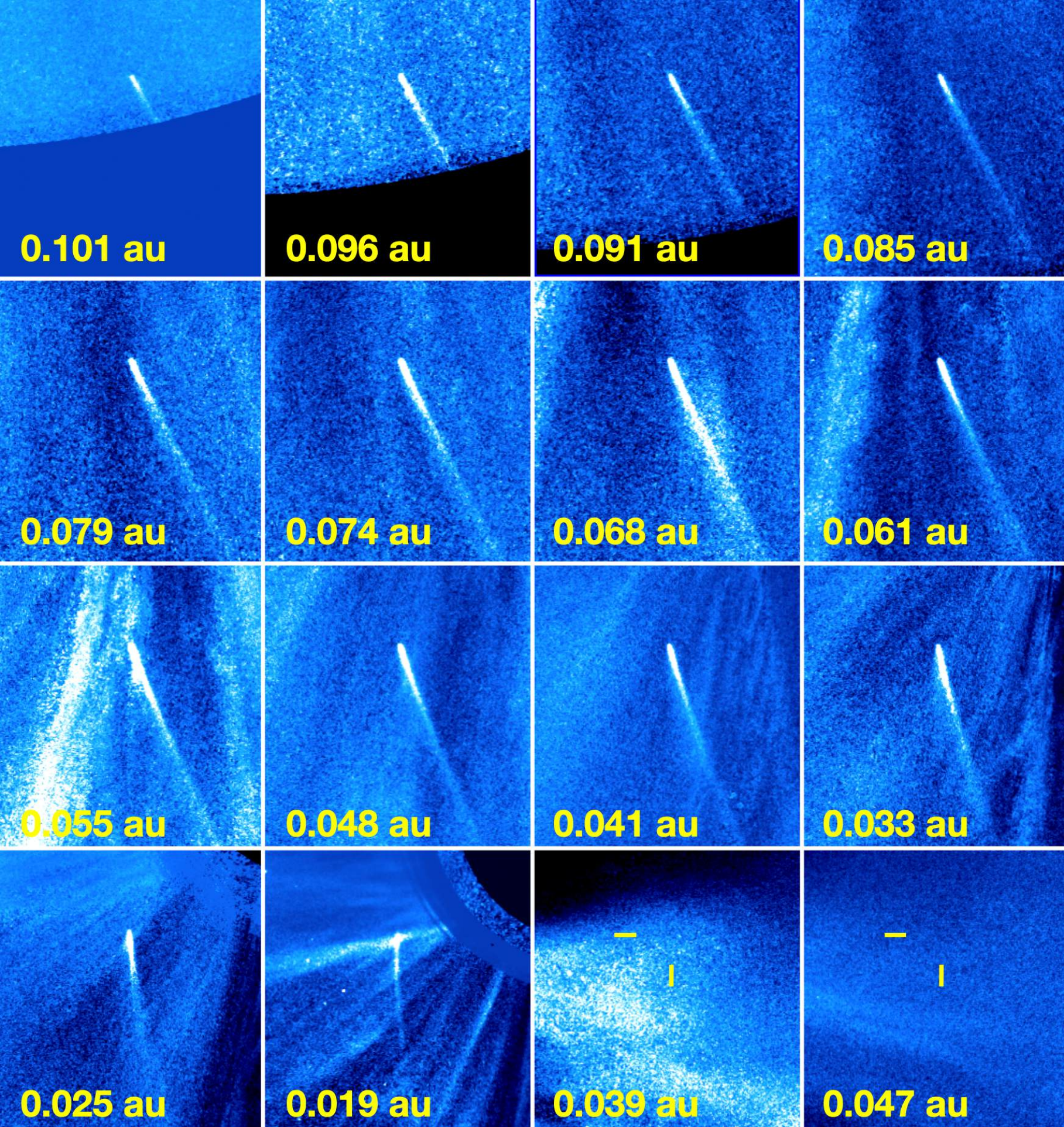}
\caption{Cor-2A composite. Each panel has north to the top, east to the left, and is approximately 1\degr~wide (2.4$\times10^6$ km at the comet). Yellow tick marks in the 0.039 au and 0.047 au panels show the expected post-perihelion location of the comet, which was not seen at these distances.  \label{Cor2}}
\end{figure}

\textbf{STEREO COR-2A Data:}  In addition to data from the NOT, we used images of S1 taken by the Sun-observing NASA satellite STEREO-A (\cite{How08}, \cite{Eyl09}).  
We used data from the COR-2A camera, which is an externally occulted Lyot coronagraph with an unobstructed field extending from $\sim$2 to 15 R$_{\odot}$.  The COR-2A camera has an image scale 14.7\arcsec~pixel$^{-1}$ (i.e., each pixel subtends a solid angle $\sim$4700 times larger than in the ALFOSC data) and an effective wavelength $\sim$7000\AA, with a 1000\AA~wide bandpass.  Images from Cor-2A are affected by the strong, highly structured and variable coronal background.  To suppress the corona then we first found the minimum brightness value in each pixel from the 114 COR-2A images taken on October 27 to 28 and subtracted this image from the individual images.  We then combined COR-2A images of 6 s integration in median groups of typically 5 to 6 images to provide better signal-to-noise ratio data.  Alignment of the images was made using field stars and the JPL Horizons ephemeris for S1.  We also used sidereally-aligned composite images of field stars, treated in the same way as the comet data, for photometric calibration. The comet entered the COR-2A field on October 27 and remained visible for approximately 24 hours.   A journal of STEREO-A observations is given in Table \ref{STEREO} and the combined images are shown in Figure \ref{Cor2}.

The spacecraft to comet distance barely changed during the $\sim$1 day over which STEREO-A observations were obtained.  Therefore, we used circular projected apertures of fixed angular radii 5 and 10 pixels (74\arcsec~and 147\arcsec), with background subtraction from the median signal computed within a contiguous annulus having outer radius 50 pixels (735\arcsec).  The 5 and 10 pixel aperture radii correspond to 48,000 km and 96,000 km, respectively, at the comet. We refer to magnitudes from these apertures as $R_{48}$ and $R_{96}$.   We checked to be sure that the use of the median signal in the sky aperture successfully removed light from the tail of the comet and did a good job in representing the residual coronal background.

The photometry was corrected to find the absolute magnitudes within the 5 and 10 pixel apertures, $H_{48}$ and $H_{96}$, respectively, as described above (Table \ref{STEREO_photometry}). Figure \ref{STEREO_plot} shows that the comet reached peak absolute magnitude at about $r_H$ = 0.075 au (16 R$_{\odot}$).  It faded thereafter by about 2 magnitudes, until the last detection at $r_H$ = 0.019 au (4.1 R$_{\odot}$), just 2.3 hours before the perihelion at 0.008 au (1.7 R$_{\odot}$).

We also examined post-perihelion COR-2A image composites in search of surviving remnants of S1.  Images acquired up to UT 2024 October 29 02h:53m ($r_H$ = 0.074 au or 16 R$_{\odot}$) show no evidence for S1, after which the ephemeris position of the comet fell outside the field of view of the coronagraph.

\section{DISCUSSION}

The observations from NOT and STEREO-A show that

1) The pre-perihelion absolute magnitude varied strongly in NOT data from the period October 3 to October 11 ($r_H$ = 0.937 au to 0.460 au) but did not follow any simple power-law type dependence, at first fading then suddenly brightening.

2) Fragments imaged on October 11 ($r_H$ = 0.725 au; Figure \ref{october11}) show that the nucleus was breaking up long before perihelion.

3) The pre-perihelion nucleus radius was not larger than $r_n$ = 600 m (albedo 0.04 assumed)

4) Peak absolute brightness in the STEREO-A data was reached on October 27 when at $r_H$ = 0.075 au ($\sim$16 R$_{\odot}$).  Thereafter, the comet faded until the last pre-perihelion observation on October 28, at $r_H$ = 0.02 au (4.1 R$_{\odot}$). 

5) The comet was not detected after perihelion, either in STEREO A observations or in more sensitive ground-based data from NOT (Table \ref{geometry}), with an upper limit to the size of a surviving compact nucleus from post-perihelion NOT data, $r_n \le$ 700 m (albedo 0.04 assumed).

\subsection{Physical Processes}

We briefly consider the  physical processes likely to influence the changing appearance and photometry of C/2024 S1.

\begin{figure}
\epsscale{0.8}
\plotone{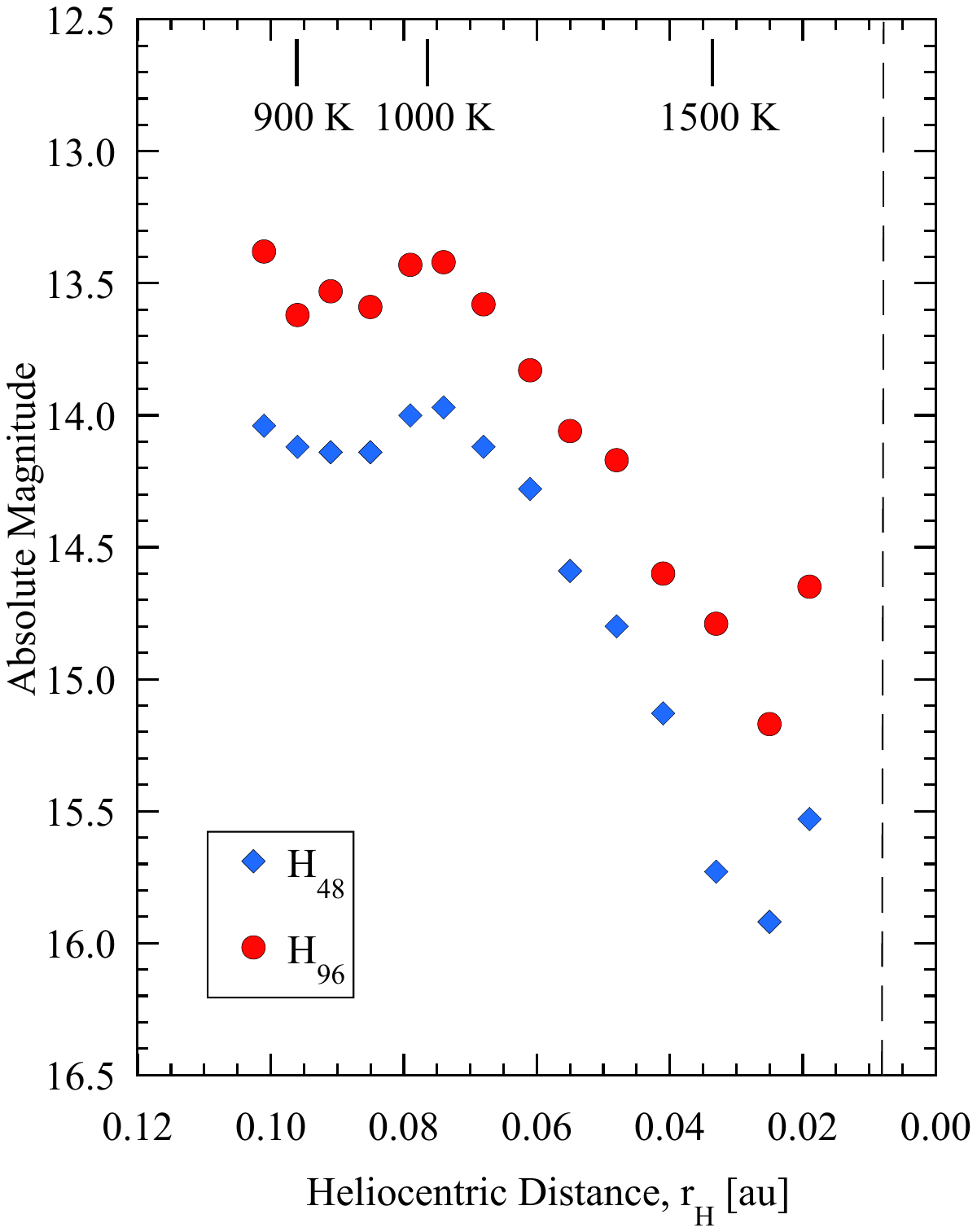}
\caption{Photometry from the COR-2A data as a function of the heliocentric distance.  The absolute magnitudes are shown within 48,000 km (blue diamonds) and 96,000 km  (red circles) radius apertures. Isothermal, spherical blackbody temperatures are marked along the top axis.  The vertical dashed line marks perihelion.\label{STEREO_plot}}
\end{figure}

\textbf{Rock Sublimation:} The equilibrium temperature of a spherical, isothermal blackbody, $T_{BB} = 278r_H^{-1/2}$, is shown along the top axis of Figure \ref{STEREO_plot}.  Comet S1 reached peak brightness near $r_H \sim$  0.075 au (16 R$_{\odot}$), corresponding to $T_{BB} \sim$ 1000 K.  Small grains (those with radius less than the wavelength of the blackbody peak) can rise to even higher equilibrium temperatures as a result of their reduced emissivity.  Temperatures in excess of 1000 K are far too high for exposed water ice to survive but are consistent with the sublimation of a more refractory material. As an example,  olivine is known to be abundant in the comae of comets \citep{Zol24} and olivine has a sublimation distance, 0.07 au ($\sim$15 R$_{\odot}$) \citep{Kob11}, very similar to the location of the brightness peak. Sublimating coma grains of olivine, or of another material with similar latent heat of sublimation, plausibly account for the near-perihelion fading of S1 seen in Figure \ref{STEREO_plot}.  Sungrazing comets in general reach peak apparent magnitudes when near $\sim$10 to 14 R$_{\odot}$ \citep{Kni10}. This has been interpreted as representing the complete dissolution of the nucleus, but sublimation of coma grains offers an additional and, indeed, unavoidable explanation.  We note that some $\beta$ meteoroids originate by  sublimation of Zodiacal dust grains in the 10 to 20 R$_{\odot}$ heliocentric distance range \citep{Sza21}.

It is important to note that the nucleus of S1 presents only a tiny fraction ($\lesssim$0.1\%) of the total (coma dominated) scattering cross-section.  Therefore, we cannot conclude from the near-perihelion fading of the comet that the nucleus was destroyed, although neither can we rule it out.

\textbf{Conduction:} Thermal conduction is not important in the destruction of the nucleus of S1.  To see this, note that in elapsed time $\tau$, heat conducts over a distance $d \sim (\kappa \tau)^{1/2}$, so that the effective speed of the conduction front, $v_c = d/\tau$ is given by $v_c \sim (\kappa/\tau)^{1/2}$.  Here, $\kappa$ is the thermal diffusivity which, for porous materials found in comets, takes the value $\kappa \sim 10^{-9}$ m$^2$ s$^{-1}$.  In the $\tau \sim$ 150 days (1.3$\times10^7$ s) between crossing the water sublimation line at 3 au and perihelion, we compute $d \sim$ 0.1 m, and the mean conduction front speed is $v_c \sim 8\times10^{-9}$ m s$^{-1}$.  The loss of ice from such a thin surface layer could have no consequence for the stability or survival of the nucleus as a whole.

\textbf{Ice Sublimation:} Mass loss from S1 is instead dominated by  ablation of the surface due to sublimation.  At distances $r_H \lesssim$ 1 au, the equilibrium sublimation rate from a perfectly absorbing, flat water ice surface exposed perpendicular to the Sun is given by energy balance as

\begin{equation}
f_s \sim \frac{L_{\odot}}{4\pi r_{H}^2 H}
\label{fs1}
\end{equation}

\noindent where $L_{\odot}$ = 4$\times10^{26}$ W is the Solar luminosity, $r_{H}$ is the heliocentric distance, and $H = 2.8\times10^6$ J kg$^{-1}$ is the latent heat of sublimation.  For example, at distances $r_{H}$ = 1.0 au (c.f.~Table \ref{geometry}), Equation \ref{fs1} gives 
$f_s \sim 0.5  \times10^{-3}$ kg m$^{-2}$ s$^{-1}$.  With bulk density $\rho$ = 500 kg m$^{-3}$ (the nominal  cometary nucleus density according to \cite{Gro19}), the speed with which the ice surface recedes  by sublimation, $v_s = f_s/\rho$, is $v_s \sim 1  \times 10^{-6}$ m $s^{-1}$, about two orders of magnitude faster than the speed of the conduction front. For this reason, sublimation from surface ice on S1 should continuously expose cold ice from beneath.

To examine this in more detail, we calculate the cumulative loss of ice as S1 moves from large heliocentric distances to perihelion.  We solved

\begin{equation}
\Delta Z(r_H) = \frac{1}{\rho} \int_{t_0}^{t} f_s(r_H(t)) dt,
\label{thickness}
\end{equation}

\noindent where $\Delta Z(r_H)$ is the thickness of an ice layer lost in moving from aphelion to distance $r_H$, $t_0$ is the time at aphelion and $t$ is the time when the comet has reached  $r_H$.  We obtained the relation between $r_H$ and $t$ by numerical solution of Kepler's equation and computed $f_s$ from the sublimation equilibrium equation

\begin{equation}
\frac{L_{\odot} \gamma(r_H)(1-A)}{4\pi r_H^2} = \chi \left(\epsilon \sigma T^4 + f_s(r_H(t)) H \right)
\label{sublimation}
\end{equation}

\noindent where $A$ and $\epsilon$ are the Bond albedo and infrared emissivity of the ice.  Neither $A$ nor $\epsilon$ is known, but their influence on the solution for $f_s$ is negligible provided $A \ll$ 1 and $\epsilon \sim$ 1.  We assumed $A$ = 0 and $\epsilon$ = 1. We included in the energy balance calculation a geometric correction, $\gamma(r_H) = 1 - (r_{\odot}/r_H)^2$, to account for the fact that an entire solar hemisphere is not visible from S1 when $r_H/R_{\odot}$ is small. (This geometric correction ignores the effects of limb darkening and fails as $r_H/r_{\odot} \rightarrow 1$, so must be regarded as approximate.  However, the effect on $\Delta Z$ is in any case small, because the comet spends very little time in close proximity to the Sun). The Stefan-Boltzmann constant is $\sigma = 5.67\times10^{-8}$ W m$^{-2}$ K$^{-4}$. Quantity $\chi$ is a dimensionless parameter that reflects the spatial distribution of heat over the surface of the nucleus.  For a flat surface oriented perpendicular to the Sun-comet line, $\chi$ = 1 while, for a spherical nucleus heated only on the day-side, $\chi$ = 2 and for an isothermal sphere, $\chi$ = 4.   Since comets are known to sublimate primarily from their hot, Sun-facing sides, we ignore the latter limit and consider only $\chi$ = 1 and 2.  

\begin{figure}
\epsscale{0.8}
\plotone{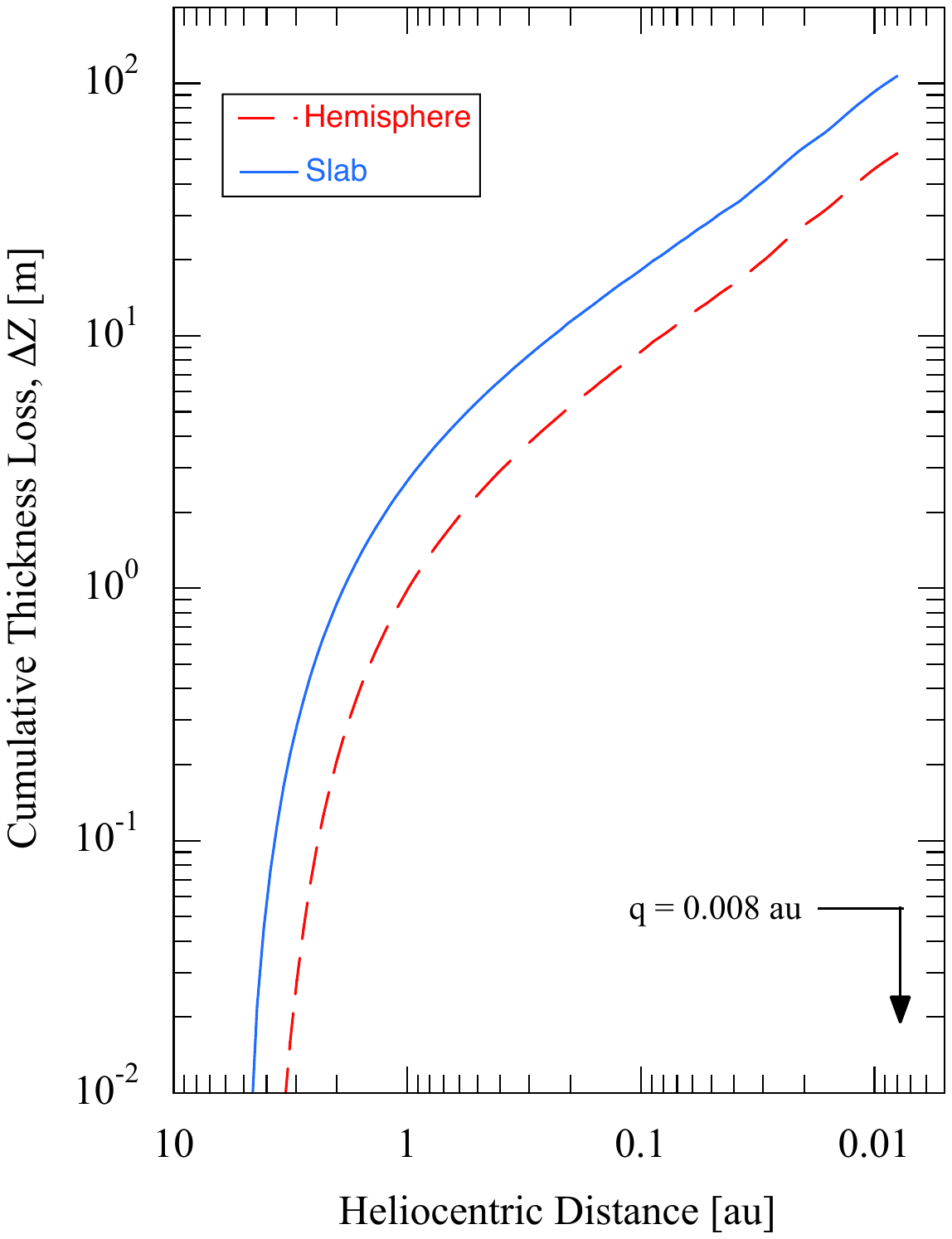}
\caption{Cumulative ice ablation as a function of decreasing heliocentric distance for two sublimation models (solid blue line) a slab oriented perpendicular to the Sun-comet line and (dashed red line) a spherical ice surface sublimating only from the Sun-facing hemisphere.  The orbit of C/2024 S1 and ice density $\rho$ = 500 kg m$^{-3}$ are assumed. \label{cumu_loss}}
\end{figure}

Equation \ref{sublimation} was solved iteratively for the mass sublimation flux of water ice, $f_s(r_H)$.  
The results, plotted in Figure \ref{cumu_loss}, show  two illumination cases.  The solid blue line shows the solution for $\chi$ = 1.  It represents the maximum possible temperature and sublimation rate at a given $r_H$, roughly corresponding to the high temperature of the subsolar point on S1.  The dashed red curve corresponds to $\chi$ = 2.  The curves are similar, and consistent with  estimates based on Equation \ref{fs1}.   Figure \ref{cumu_loss} shows that the cumulative loss of ice from aphelion down to 1 au  is $\Delta Z \sim$ a few meters, while extended down to perihelion the loss approaches $\Delta Z \sim$ 50 m to 100 m.  Uninhibited sublimation on the way to perihelion could destroy a nucleus smaller than 50 m to 100 m in radius and is a plausible (but not unique - see below) cause of the demise of most Kreutz comets, these being generally smaller \citep{Kni10}.  Our calculation of $\Delta Z$ is clearly idealized but it serves to show the considerable thermal consequence of the small perihelion distance.

\textbf{Rotational Instability:} Anisotropic loss of mass from a sublimating nucleus exerts a torque  that can change the spin.  Given enough time, this torque can spin up the rotation to the point of instability.  Recent measurements of short-period comets show that  the outflow momentum is mostly radial to the nucleus, but that a median fraction, $k_T \sim$ 0.7\%,  torques the nucleus \citep{Jew21}.   Considering the nucleus to be a sphere of initial radius $r_n$, density $\rho$, in rotation at angular frequency $\omega$, the spin angular momentum may be written $L = (2/5) M_n r_n^2 \omega$, where $M_n = 4 \pi \rho r_n^3/3$ is the nucleus mass, giving

\begin{equation}
L = \frac{8\pi}{15} \rho r_n^5 \omega.
\end{equation}

\noindent The mass of a  shell of thickness $\Delta r$ is $\Delta m = 4\pi r_n^2 \rho \Delta r$, its linear momentum  when ejected at speed $V_{th}$ is $\Delta m V_{th}$ and the change in the angular momentum due to the ejection of the shell is $\Delta L = \Delta m V_{th} k_T r_n$, or 

\begin{equation}
\Delta L = 4 \pi r_n^3 \rho \Delta r V_{th} k_T.
\end{equation}

\noindent Setting $\Delta L = L$ gives an expression for $\Delta r$, the  thickness of a shell that must be lost in order to substantially change the angular momentum of the nucleus, 

\begin{equation}
\Delta r = \frac{2 r_n^2 \omega}{15 k_T V_{th}}.
\label{Deltar}
\end{equation}

To evaluate $\Delta r$ we set $r_n$ = 600 m, the maximum value allowed by the pre-perihelion photometry.  We take the coma expansion speed at $r_H$ = 1 au as $V_{th} \sim $ 500 m s$^{-1}$, equal to the thermal speed of water molecules at 300 K. The angular frequency of the nucleus of S1 is not known but the  rotation periods of most cometary nuclei are within a factor of two of $\sim$10 hours ($\omega \sim 1.7\times10^{-4}$ s$^{-1}$) \citep{Kok17}. Substituting, we find that a 600 m radius nucleus need sublimate a layer only a modest  $\Delta r$ = 2.3 m thick  in order to change its rotation by factor of order two.  Figure \ref{cumu_loss} shows that cumulative sublimation losses reach 2.3 m for inbound heliocentric distances $r_H$ = 0.5 au to 1.1 au (depending on the spatial distribution of surface temperature), consistent with the $\sim$0.7 au distance at which S1 first showed a fragmented appearance (Figure \ref{october11}) and a dramatic surge in brightness (Figure \ref{NOT_plot}).  While this is not proof that the nucleus reached rotational instability at these pre-perihelion distances, this inference is certainly plausible provided the nucleus is not structurally cohesive.   

Equation \ref{Deltar} shows that $\Delta r$ scales as $r_n^2$ so that, if the nucleus is smaller than 600 m in radius, the torque due to ejection of an even thinner surface layer would be sufficient to spin-up the nucleus and disintegration could start at even larger heliocentric distances.  Most Kreutz comets are small (a few to $\sim$50 m) and do not survive to perihelion.  Following Equation \ref{Deltar}, a 50 m body need lose only $\sim$2 cm in order for outgassing torques to materially change its spin.  Figure \ref{cumu_loss} shows that ice loss of 2 cm is possible even at asteroid belt distances, suggesting that breakup can begin long before most small Kreutz comets are discovered.  As with asteroids, where rapid rotation is largely confined to $<$100 m bodies \citep{War09},  survival against rotational disintegration  may reflect the role of finite cohesive strength.    

At the other extreme, Equation 8 indicates that very large Kreutz comet nuclei should survive against rotational disruption given the limited opportunity to sublimate on the way to perihelion. Substituting $\Delta r$ = 50 m to 100 m (the maximum cumulative sublimation loss measured down to perihelion; see Figure 6) into Equation 8, with the other parameters as above, gives the radius of the largest nucleus that can be spun up on the way to perihelion as $r_n$ = 2.8 km to 3.9 km. Unfortunately, the dimensions of the largest Kreutz comet nuclei are not observationally well-determined. Sungrazer C/2011 W3 (Lovejoy) evidently disintegrated near perihelion; its nucleus size, while uncertain, was almost certainly less than 1 km \citep{Ray18}.  Table 1 of Knight et al. (2010) lists the (also very uncertain) radius of perihelion survivor C/1965 S1 (Ikeya-Seki) as 4.3 kilometers. Large Kreutz comets can still be split by tidal forces, as was C/1965 S1 \citep{Sek66}, but their disruption by rotational bursting is less likely.

These considerations are clearly very crude; the nucleus is probably not spherical, mass is probably not lost uniformly from a surface shell,  the  median value of $k_T$ measured in short-period comets  \citep{Jew21} might not apply to S1 and the comets, although structurally weak, might have non-negligible strength with which to resist rotational bursting. Nevertheless, spin-up from outgassing torques is a plausible cause of breakup in this and other sub-kilometer sungrazing comets and one that offers a fast path to nucleus destruction.  Since only a few meters of surface must be lost in order to drive the nucleus towards breakup, rotational disruption is the most efficient route leading to nucleus destruction.

\textbf{Roche Instability:}
The gradient of solar gravity across the nucleus diameter is a potential cause of disruption inside the Roche distance

\begin{equation}
\frac{r_R}{R_{\odot}} \sim \Gamma \left(\frac{\rho_{\odot}}{\rho}\right)^{1/3}
\label{roche}
\end{equation}

\noindent where $\Gamma$ is a dimensionless constant and $\rho_{\odot}$ = 1300 kg m$^{-3}$ is the density of the Sun.  For a strengthless sphere, $\Gamma \sim$ 2.4, and with $\rho$ = 500 kg m$^{-3}$ Equation \ref{roche} gives $r_R/R_{\odot} \sim$ 3.3. This is only approximate because the nucleus is unlikely to be either strengthless or spherical and because Equation \ref{roche} further neglects the destabilizing influence of nucleus rotation.  The instability distance could be either larger or smaller than $r_R/R_{\odot} \sim$ 3.3 depending on its strength, shape and rotation. Nevertheless, it is evident that tidal disruption near perihelion (where $r_H/R_{\odot} \sim$ 1.7) is a plausible contributor to the apparent demise of S1, just as was the case for  comet D/1993 F2 (Shoemaker-Levy 9), which split upon entering the Roche sphere of Jupiter \citep{Asp96}, and C/1965 S1 (Ikeya-Seki), which split at its 0.008 au perihelion \citep{Sek66}.

Comet S1 is unusual in that it was bright enough to be discovered at $\sim$1 au, while most Kreutz comets are too small (radii from a few to $\sim$50 m \citep{Kni10}) and too faint to be detected at this distance.   The early detection of S1  suggests that its nucleus is larger than average, with our empirical upper limit being $r_n <$ 600 m.   A few even larger Kreutz comets are known. For example, the nucleus of C/2012 S1 (ISON) was slightly larger at 600 $\le r_n \le$ 900 m  \citep{Lam14} and the most famous Kreutz comet, C/1965 S1 (Ikeya-Seki) was probably $>$1 km in radius, although reliable nucleus size estimates for this comet do not exist. Most Kreutz comets vanish before or near perihelion, as did C/2024 S1 and C/2012 S1 (ISON) \citep{Kea16}.   Only sungrazers large enough to survive rotational and ablative destruction face the final gauntlet of gravitational disruption (Equation \ref{roche}) upon entering the Sun's Roche sphere.

\clearpage

\section{SUMMARY}
Optical observations of Kreutz sungrazing comet C/2024 S1 from ground- and space-based telescopes show that 

\begin{enumerate}

\item The pre-perihelion nucleus radius was $r_n <$ 600 m (red geometric albedo 0.04 and C-type phase function assumed).  Disintegration of the nucleus into multiple components was underway by 0.7 au pre-perihelion.

\item The geometry-corrected ground-based photometry shows no simple dependence on heliocentric distance, with the comet at first fading by $\sim$1 magnitude from 0.9 au to 0.7 au, then brightening by 3.5 magnitudes towards 0.45 au, indicating non-equilibrium mass loss.

\item Near-Sun photometry from STEREO-A shows that S1 reached peak brightness near 0.075 au, where the radiative temperature is $\sim$1000 K, and faded on the way to perihelion.  Grain sublimation may explain this near-perihelion fading.

\item Rotational instability caused by outgassing torques is the most efficient mechanism of destruction for this and other Kreutz sungrazing comets, followed by ablative losses due to sublimation.  Nucleus fragments surviving rotational disruption and ablation on the way to perihelion would have been disrupted upon entering the Sun's Roche sphere.

\end{enumerate}

\acknowledgments
We thank two anonymous referees for their comments.  The ground-based data presented here were obtained with ALFOSC, which is provided by the Instituto de Astrofisica de Andalucia (IAA) under a joint agreement with the University of Copenhagen and NOT.





\begin{deluxetable}{lcccrrrrrcl}
\tabletypesize{\scriptsize}
\tablecaption{Journal of NOT Observations 
\label{geometry}}
\tablewidth{0pt}
\tablehead{\colhead{UT Date\tablenotemark{a}} & \colhead{DOY$_{24}$\tablenotemark{b}}  & \colhead{$r_H$\tablenotemark{c}} & \colhead{$\Delta$\tablenotemark{d}}  & \colhead{$\alpha$\tablenotemark{e}} &  \colhead{$\varepsilon$\tablenotemark{f}} & \colhead{$\theta_{- \odot}$\tablenotemark{g}} & \colhead{$\theta_{-V}$\tablenotemark{h}} & \colhead{$\delta_{\oplus}$\tablenotemark{i}} & \colhead{Instr\tablenotemark{j}}  & \colhead{Seeing\tablenotemark{k}} }

\startdata
2024 Oct 03 & 277 & 0.937 & 1.271 & 51.2 & 46.8 & 262.8 & 264.5 & 12.1 & A & 6\arcsec \\
2023 Oct 04 & 278 & 0.911 & 1.244 & 52.6 & 46.3 & 262.8 & 264.6 & 12.9 & A & 2\arcsec\\
2024 Oct 05 & 279 & 0.886 & 1.217 & 54.1 & 45.8 & 262.8 & 264.7 & 13.6 & A & $<$1\arcsec\\
2024 Oct 06 & 280 & 0.860 & 1.190 & 55.6 & 45.2 & 262.8 & 264.7 & 14.5 & A &  1.3\arcsec\\
2024 Oct 09 & 283 & 0.780 & 1.112 & 60.7 & 43.0 & 262.3 & 264.5 & 17.1 & A & 1.0\arcsec\\
2024 Oct 10 & 284 & 0.753 & 1.088 & 62.6 & 42.0 & 262.2 & 264.3 & 18.0 & A & 1.0\arcsec\\
2024 Oct 11 & 285 & 0.725 & 1.064 & 64.6 & 41.0 & 261.8 & 264.1 & 19.0 & A & $<$1\arcsec\\
2024 Oct 17 & 291 & 0.544 & 0.941 & 79.4 & 32.5 & 258.2 & 261.1 & 26.1 & S & 1\arcsec\\
2024 Oct 19 & 293 & 0.476 & 0.911 & 85.7 & 28.5 & 256.1 & 259.3 & 28.5 & A & 1\arcsec \\
2024 Oct 20 & 294 & 0.440 & 0.899 & 89.2 & 26.2 & 254.9 & 258.3 & 29.7 & A & 2\arcsec \\
2024 Nov 06 & 311 & 0.460 & 1.000 & 75.6 & 26.7 & 265.4 &  81.4 & 27.9 & A & 2\arcsec \\
2024 Nov 10 & 315 & 0.593 & 0.992 & 72.4 & 34.8 & 266.9 &  83.1 & 30.0 & A & 1.4\arcsec \\
\enddata


\tablenotetext{a}{UT Date of observations}
\tablenotetext{b}{Day of Year.  1= UT 2024 January 1}
\tablenotetext{c}{Heliocentric distance, in au }
\tablenotetext{d}{Geocentric distance, in au }
\tablenotetext{e}{Phase angle, in degrees }
\tablenotetext{f}{Elongation angle, in degrees }
\tablenotetext{g}{Position angle of projected anti-solar direction, in degrees }
\tablenotetext{h}{Position angle of negative heliocentric velocity vector, in degrees}
\tablenotetext{i}{Angle of observatory from orbital plane, in degrees}
\tablenotetext{j}{Detector A = ALFOSC, S = StanCAM}
\tablenotetext{k}{Seeing full width at half maximum}

\end{deluxetable}

\clearpage

\begin{deluxetable}{lcccrrrrrcl}
\tablecaption{Journal of STEREO A Observations 
\label{STEREO}}
\tablewidth{0pt}
\tablehead{\colhead{UT 2024 Oct\tablenotemark{a}} & \colhead{DOY$_{24}$\tablenotemark{b}}  & \colhead{$r_H$\tablenotemark{c}} & \colhead{$\Delta$\tablenotemark{d}}  & \colhead{$\alpha$\tablenotemark{e}} &  \colhead{$\varepsilon$\tablenotemark{f}} & \colhead{$\theta_{- \odot}$\tablenotemark{g}} & \colhead{$\theta_{-V}$\tablenotemark{h}} & \colhead{$\delta_{\oplus}$\tablenotemark{i}} }

\startdata
27d 12h 09m & 301.506 & 0.101 & 0.886 & 138.9 & 4.0 & 193.1 & 215.5 & 43.5\\
27d 14h 00m & 301.580 & 0.096 & 0.891 & 139.3 & 3.7 & 191.3 & 214.7 & 43.2 \\
27d 15h 38m & 301.652 & 0.091 & 0.894 & 139.7 & 3.5 & 189.5 & 213.9 & 43.0 \\
27d 17h 18m & 301.721 & 0.085 & 0.898 & 140.1 & 3.2 & 187.4 & 213.0 & 42.8 \\
27d 19h 00m & 301.791 & 0.079 & 0.903 & 140.4 & 3.0 & 185.1 & 212.0 & 42.5 \\
27d 20h 39m & 301.860 & 0.074 & 0.907 & 140.7 & 2.8 & 182.5 & 210.9 & 42.2 \\
27d 22h 18m & 301.929 & 0.068 & 0.912 & 141.0 & 2.5 & 179.4 & 209.6 & 41.9\\
28d 00h 00m & 302.000 & 0.061 & 0.917 & 141.1 & 2.3 & 175.7 & 208.0 & 41.5 \\
28d 01h 32m & 302.064 & 0.055 & 0.921 & 141.1 & 2.1 & 171.6 & 206.2 & 41.1 \\
28d 03h 18m & 302.137 & 0.048 & 0.927 & 140.8 & 1.8 & 165.9 & 203.5 & 40.5\\
28d 05h 00m & 302.208 & 0.041 & 0.934 & 139.9 & 1.6 & 158.7 & 200.1 & 39.7 \\
28d 06h 38m & 302.227 & 0.033 & 0.940 & 138.0 & 1.3 & 149.4 & 195.1 & 38.5 \\
28d 08h 17m & 302.346 & 0.025 & 0.948 & 133.2 & 1.1 & 136.1 & 186.7 & 36.3 \\
28d 09h 24m & 302.391 & 0.019 & 0.954 & 125.7 & 0.9 & 123.2 & 176.4 & 33.4 \\
28d 18h 09m & 302.756 & 0.039 & 0.967 &  85.8 & 2.3 & 272.5 &  76.1 & 29.0 \\
28d 20h 00m & 302.833 & 0.047 & 0.965 &  89.5 & 2.8 & 269.6 &  74.6 & 29.7 \\
\enddata


\tablenotetext{a}{UT Date of observations}
\tablenotetext{b}{Day of Year. 1= UT 2024 January 1}
\tablenotetext{c}{Heliocentric distance, in au }
\tablenotetext{d}{Spacecraft to comet distance, in au }
\tablenotetext{e}{Phase angle, in degrees }
\tablenotetext{f}{Elongation angle, in degrees }
\tablenotetext{g}{Position angle of projected anti-solar direction, in degrees }
\tablenotetext{h}{Position angle of negative heliocentric velocity vector, in degrees}
\tablenotetext{i}{Angle of observatory from orbital plane, in degrees}

\end{deluxetable}

\clearpage

\begin{deluxetable}{clll}
\tablecaption{NOT Photometry\tablenotemark{a} 
\label{NOT_photometry}}
\tablewidth{0pt}
\tablehead{\colhead{UT Date} & \colhead{R$_{5}$/H$_5$/C$_5$}  & \colhead{R$_{10}$/H$_{10}$/C$_{10}$} & \colhead{R$_{15}$/H$_{15}$/C$_{15}$}   }

\startdata

3	&	16.12	/	14.57	/	36	&	15.52	/	13.97	/	63	&	15.20	/	13.65	/	84	\\
4	&	15.66	/	14.22	/	50	&	15.00	/	13.56	/	92	&	14.72	/	13.28	/	119	\\
5	&	15.72	/	14.38	/	43	&	15.05	/	13.71	/	80	&	14.74	/	13.40	/	106	\\
6	&	15.78	/	14.56	/	36	&	15.11	/	13.89	/	68	&	14.80	/	13.58	/	90	\\
9	&	15.75	/	14.90	/	26	&	15.01	/	14.16	/	52	&	14.71	/	13.86	/	69	\\
10	&	15.83	/	15.12	/	21	&	15.16	/	14.45	/	40	&	14.85	/	14.14	/	54	\\
11	&	15.90	/	15.34	/	17	&	15.23	/	14.67	/	33	&	14.92	/	14.36	/	44	\\
17	&	13.93	/	14.48	/	39	&	13.46	/	14.01	/	61	&	13.35	/	13.90	/	67	\\
19	&	10.02	/	11.06	/	922	&	9.52	/	10.56	/	1460      &	9.33	/	10.37	/	1741	\\
20	&	9.90	/	11.21	/	801	&	9.39	/	10.70	/	1280	&	9.21	/	10.52	/	1512	\\
\enddata


\tablenotetext{a}{Photometry listed in the format apparent/absolute/cross-section [km$^2$] for the 5000 km, 10,000 km and 15,000 km radius apertures.}
\end{deluxetable}

\clearpage

\begin{deluxetable}{lll}
\tablecaption{STEREO Photometry\tablenotemark{a} 
\label{STEREO_photometry}}
\tablewidth{0pt}
\tablehead{\colhead{UT Date} & \colhead{R$_{48}$/H$_{48}$/C$_{48}$}  & \colhead{R$_{96}$/H$_{96}$/C$_{96}$}    }

\startdata
301.510	&	7.27	/	14.04	/	59	&	6.61	/	13.38	/	109	\\
301.580	&	7.23	/	14.12	/	55	&	6.73	/	13.62	/	87	\\
301.650	&	7.11	/	14.14	/	54	&	6.50	/	13.53	/	95	\\
301.720	&	6.95	/	14.14	/	54	&	6.40	/	13.59	/	89	\\
301.790	&	6.64	/	14.00	/	61	&	6.07	/	13.43	/	104	\\
301.860	&	6.46	/	13.97	/	63	&	5.91	/	13.42	/	105	\\
301.930	&	6.41	/	14.12	/	55	&	5.87	/	13.58	/	91	\\
302.000	&	6.34	/	14.28	/	48	&	5.89	/	13.83	/	72	\\
302.060	&	6.44	/	14.59	/	36	&	5.91	/	14.06	/	58	\\
302.140	&	6.39	/	14.80	/	29	&	5.76	/	14.17	/	52	\\
302.210	&	6.45	/	15.13	/	22	&	5.92	/	14.60	/	35	\\
302.230	&	6.72	/	15.73	/	13	&	5.78	/	14.79	/	30	\\
302.350	&	6.66	/	15.92	/	11	&	5.91	/	15.17	/	21	\\
302.390	&	6.16	/	15.53	/	15	&	5.28	/	14.65	/	34	\\
\enddata


\tablenotetext{a}{Photometry listed in the format apparent/absolute/cross-section  [km$^2$] for the 48,000 km and 96,000 km COR-2A photometry apertures.}
\end{deluxetable}



\clearpage

%
%

%
%

\clearpage

\clearpage

\clearpage

\clearpage 

\clearpage

%

\end{document}